\documentclass[twocolumn]{article}
\usepackage{amsmath,amssymb}
\def\lambdabar{{\mathchar'26\mkern-10mu\lambda}}
\input xy
\xyoption{all}
\xyoption{arc}
\usepackage[curve]{xy}
\xyoption{poly}

\begin{document}

\title{Physical Interpretations of\\
Relativity Theory Conference IX\\
London, Imperial College, September, 2004\\
\dots\dots\dots\dots\dots\\
Mach's Principle II}
\author{by\\
   \\
James G. Gilson\thanks{j.g.gilson@qmul.ac.uk}\\
School of Mathematical Sciences\\
Queen Mary College\\
University of London}

\maketitle

\begin{abstract}
The meaning and significance of {\it Mach's Principle\/} and its dependence on ideas about relativistic rotating frame theory and the celestial sphere is explained and discussed.  Two new relativistic rotation transformations are introduced by using a linear simulation for the rotating disc situation. The accepted formula for centrifugal acceleration in general relativity is then analysed with the use of one of these transformations. It is shown that for this general relativity formula to be valid throughout all space-time there has to be everywhere a local standard of absolutely zero rotation. It is then concluded that the field off all possible space-time null geodesics or photon paths unify the absolute local non-rotation standard throughout space-time. Thus it is suggested that Mach's principle holds in the restricted sense that 
{\it there is a universal standard of absolute local rotation rate related to the apparent rotation of the celestial sphere\/}. However this apparent rotation is actually the earth's rotation relative to a local mapping of null geodesic endpoints from that time and space distant sphere to the local time in the local zero-rotation environment. A connection of local inertia with the celestial sphere is not found.
  
\end{abstract}
\vskip 0.2cm

\section{Introduction}
\setcounter{equation}{0}
\label{sec-intro}

The question of the validity or otherwise of Mach's Principle\cite{Mac:1893}   has been with us for almost a century and has been vigorously analysed and discussed for all of that time and is, up to date, still not resolved. One contentious issue is, does general\cite{Mis:1973} relativity theory imply that Mach's Principle is operative in that theoretical structure or does it not? A deduction\cite{Rin:2001} from general relativity that the classical Newtonian centrifugal acceleration formula should be modified by an additional relativistic $\gamma ^2 (v)$ factor, see equations (\ref{eq:1.3}) and (\ref{eq:4.1}), will help us in the analysis of that question. Here I shall attempt to throw some light on the nature and technical basis of the collection of technical and philosophical problems generated from Mach's Principle and give some suggestions about how it can, at the very least, be clarified technically and philosophically. Firstly, let us review the physical issues involved and try to give a clear definition of exactly what the problem is. It is usually explained in terms of what is called Newton's bucket experiment\cite{Bar:1995}. A bucket containing water is spun about a vertical axis through its centre. The contained water picks up motion because of frictional drag by the bucket wall and then assumes a concave downwards surface form as a result of the outward centrifugal force causing the rise in level from the centre towards the wall. This is a very simple sequence of physical events which each of us can or has experienced in some form or other but it prompts a philosophical question which is difficult to answer. That such a fundamental physical-philosophical dilemma as exposed by this simple experiment is so immediately accessible to everyone's understanding makes it unique among scientific problems. The question is how does the water in the bucket get the information that it is rotating so as to respond with the concave surface? Certainly the wall of the bucket has caused the rotation via friction but it has not caused the centrifugal {\it force\/} response. The situation is further confused when one asks what is the water rotation relative to? Certainly there is the rotation relative to the room in which the experiment takes place but what if that room is itself rotating so slowly as not to have caused a noticeable concavity initially. Is the water responding to just that initially induced room relative rotation or to some combination of the two conceivable contribution to the water angular velocity? It is usually taken to be the case that the water rotation and the induced centrifugal force are  consequential upon a totality of any rotational motions of the body about the direction involved.

Another much discussed example and the one which seems most impressive to me is that of the influence of the rotating state on a physical elastic sphere. It is certainly true that such a sphere set into rotation will expand in girth about the equator of its rotary axis and to a degree dependent on its angular velocity. Then the question is how does it acquire the information of the magnitude of its angular velocity unless there is some local standard of zero rotation? From astronomy technology, we know that 
a telescope can be locked on to a distant galaxy for some finite time in order to photograph it. This requires mechanical and computer equipment to negate the effect of the earth's rotation so that the axis of the telescope can be kept on a fixed line direction pointing towards the galaxy so that it remains in view at a central fixed position for the finite time necessary. All such viewable galaxies are part of what is called the celestial sphere and a specific known galaxy can be used to indicate a definite coordinate position on that sphere. The mechanisms and systems required to construct such a viewing system for studying the celestial sphere is a commonplace available technology of the present day. This telescopic viewing system can be regarded as giving an identification and measuring device for the geometry of the celestial sphere. There is another way to lock a local line axis onto an object on the celestial spheres and that is by using a property of angular momentum. An extended body which is of material construction so that it is composed of some distribution of mass within its physical boundaries will acquire angular momentum if it is rotated about an axis. If further the body is not subjected to any additional external couples, the axis of rotation of the body then maintains a fixed direction with respect to the celestial sphere, with an axis of rotation aligned with respect to a specific galaxy of choice. This line axis locking system is an available gyroscopic device which is also well within existing modern technology and its equivalent is to be found in all space vehicular guidance systems. It is important to realise two things about this coupling of local angular momentum to a distant object. One is that the locking can be disturbed by applying couples to the object and the other is that the strength of the lock does depends on the magnitude of the angular momentum and only reduces to zero that is it becomes unlocked if the magnitude of the angular momentum is reduced to zero. It is this gyroscopic type of set up which is usually quoted to explain one of the aspects of {\it Mach's Principle\/}. This can be expressed as the following question. How is it that a locally rotating object can for some substantial finite length of time hook onto a distant galaxy to which it is obviously not connected? This experimental fact is the source of the idea that inertia or mass controlled influences locally are determined by the mass space fabric of the rest of the universe. Another manifestations of the mystery link between here and the distant universe should be mentioned. The Foucault Pendulum swings in a moving plane with a motion that is locked onto the apparent motion of the celestial sphere. This gives the impression that the celestial sphere determines the local state of no rotation against which the earth actually rotates. This device can be seen in museums and indeed can be easily constructed by ordinary people.

The two main answers given to these questions are what would be Newton's that there is an {\it absolute\/} space background everywhere and this defines the local angular velocity condition for any physical body. There is then what would be Mach's explanation that the angular velocity of a body anywhere should be referred to a frame of reference attached to the celestial sphere of so called fixed stars. This is a rather condensed translation of the two points of view, not exactly as expressed by either Newton or Mach. In fact, what these two researchers actually meant by their theoretic constructs is also a matter of much debate. One can see this interpretational type of difficulty  by reading the following quotation from Mach's writings given by Soshichi Uchii \cite{Sos:1980} on his website.

 {\it The comportment of terrestrial bodies with respect to the earth is 
    reducible to the comportment of the earth with respect to the remote 
    heavenly bodies. If we were to assert that we knew more of moving objects 
    than this their last-mentioned, experimentally-given comportment with 
    respect to the celestial bodies, we should render ourselves culpable of a 
    falsity. When, accordingly, we say, that a body preserves unchanged its 
    direction and velocity in space, our assertion is nothing more or less than 
    an abbreviated reference to the entire universe. (ch.2, vi-6, 285-6) 

    The considerations just presented show, that it is not necessary to refer 
    the law of inertia to a spacial absolute space. On the contrary, it is 
    perceived that the masses that in the common phraseology exert forces on 
    each other as well as those that exert none, stand with respect to 
    acceleration in quite similar relations. We may, indeed, regard all masses 
    as related to each other. That accelerations play a prominent part in the 
    relations of the masses, must be accepted as a fact of experience; which 
    does not, however, exclude attempts to elucidate this fact by a comparison 
    of it with other facts, involving the discovery of new points of view.
 (ch.2, vi-8, 288)\/}

However, in spite of these interpretation uncertainties, the description of the two points of view, as I have expressed them above, do encapsulate the essence of the problem.  The conceptual difficulties in both points of view become clearer when one tries to update them in terms of relativity theory ideas. Newton's point of view seems to fail because absolute space, as far as velocity is concerned, is ruled out and Mach's view seems untenable because it is difficult to see how objects in deep space can determine local properties. To prove which view is correct in modern relativity terms is difficult because up to date there is not a satisfactory relativity theory\cite{Wei:1988} for the rotating {\it rigid\/} body and indeed such conceptual areas are controversial and not finally settled. Another difficulty that immediately emerges when one thinks about a rotating body in relativistic terms arises from the length contraction effect\cite{Rin:2001} that is intrinsic to special relativity. For a small radius disk rotating at high angular velocity or a large radius disk rotating at small angular velocity there can be a radius on the disk that is moving transversally at the speed of light. Thus it seems that special relativity length contraction implies that the circumference at this radius will be of zero length on the basis of measurements made by a stationary observer. Geometrically one can try to explain this happening by asserting that  the originally flat disc will  have to curl up or down from centre toward edge to form a closed sphere.  However, it is difficult to imagine what such an occurrence would look like from the point of view of a non-rotating observer. From his point of view a two dimensional flat object would have become a three dimension object. This type of difficulty has led some authors\cite{Wei:1988}  to remark that the change of geometry that would seem to occur as a result of relativity theory when a flat object rotates cannot realistically physically be seen as embedded in the space of three dimensional Euclidean  geometry. Here then, is a working statement of Mach's Principle, not as expressed by Mach but, in two sentences covering the two aspects of the principle and in a form as it impinges on the subject matter of this article:-

 {\it Observers fixed in local non-rotating reference frames see the distant celestial sphere as having zero rotation relative to themselves. The inertial properties of local particles are somehow determined by all the rest of the particles of the universe with the very distant masses greatly involved.\/}

This can also be taken as the definition for such a {\it non-rotating\/} frame according to Mach.

I have found a relativistic generalization of the usual formula that is used in both classical and relativistic theory to explain the effect of a rotating body on observations from such a platform as opposed to the view from a non-rotating body. Usually the classical formula has to be adapted by some sort of ad hock addition of a second time variables to make it have some conformity with the two {\it times\/} that are involved in the Lorentz transformation between two inertial frames in linear relative motion. The inadequacy in attempts at relativistic analyses of rotating systems is of course due to a full relativistic rotation transformation formula having not to date been obtained. The relativistic rotation transformation formula that I am intending to use here will be detailed in the following pages. However, I do not wish to claim that this formula is the last word in explaining the relativistic rotating state. One reason for not making such a claim is that there appear to be a number of possibilities with slightly different physical meanings which I suspect will have applications to slightly different physical situations. So I shall take the rather cautious course of introducing this new formula within a mathematical model of the physical philosophical situation that the Mach Principle dilemma seems to be describing. This model shows that the Mach-Newton-Einstein\cite{Bel:1987} conundrum can be greatly clarified.  Hopefully it will be possible to go further and show that our physical universe conforms to the pattern of mathematical-physical quantities uncovered in this model. Whether or not Einstein's general\cite{Bar:1995} relativity conforms to this principle still seems to be an open question and indeed as we have seen it is far from clear just what the {\it Principle\/} is actually about in any precise sense. Einstein is said to have been guided by Mach's principal but it is unlikely Mach and Einstein would be able to agree on its meaning. I shall not attempt to reference the vast literary archives on this subject but for those in need of finding out more, I suggest reading the introductory article by Soshichi Uchii \cite{Sos:1980} and follow back references from there.  

In analysing the collection of problems that arise in the quest for an understanding of {\it Mach's Principle\/}, I shall here make use of a new relativistic rotation transformation, (\ref{eq:1.1} $\rightarrow$ \ref{eq:1.7}) which is one of four or more possibilities that I have found based on a simple rotation simulation idea. The transformation I place most emphasis on is the first of the pair displayed in the next section This formula can be derived rigorously from special relativity under a simple assumption about rotation. The second relativistic rotation transformation, (\ref{eq:2.1} $\rightarrow$ \ref{eq:2.9}) has a number of very attractive mathematical aspects and I suspect that it could have physical significance under some circumstance. It is not directly derivable from special relativity and will not be used in this article.
\subsection{Possible Rotation\newline Transformations}
\setcounter{equation}{0}
\label{sec-rerot}
The two relativistic rotation transformation given by equations (\ref{eq:1.1} $\rightarrow$ \ref{eq:1.7}) and equations (\ref{eq:2.1} $\rightarrow$ \ref{eq:2.9}) below are constructed with reference to a mathematical simulation process that generates the equivalent of a rotating body from the relative linear motions of  Lorentz  frames of reference $\tilde{S}$ and $\tilde{S}'$ depicted in {\it The Relativistic Rotating Frame Diagram\/} by coordinates with the top $\sim$ symbol. The construction is a follows. The basic reference frame $S$ is taken to be rectangular, rigorously inertial and to have a synchronised time keeping system throughout its space extent.  That is importantly, it is not linearly accelerating and it is not rotating. To logically continue explaining the construction we have to now introduce a space time path $P(t)$ in $S$ which will be denoted in polar coordinates by $(r(t), \theta (t), t)$ and a second frame of reference $S'$ which has a common origin with $S$ and is rotating with an angular velocity $\omega$ with respect to $S$ about that common origin. Let us now imagine a disc shaped region of the frame $S'$ centred at the common origin of $S$ and $S'$.  If the path $P(t)$ is at position $P(t)$ in the diagram at time $t$ this defines a radius of the disc $r(t)$ at time $t$ and also the circle shown in the diagram. It follows that the path $P(t)$ coincides with a point on the disc that is moving tangentially with velocity $v(t)=r(t)\omega$ anti-clockwise. Now, we can conjure up a special relativity {\it inertia\/} frame $\tilde{S}'$ moving with this velocity and in the same tangential direction as the motion at $P(t)$ which has come up to coincidence at $P(t)$ at time $t$ along a straight path from a position $L(0)$ on $S$ at time $t=0$. Let us choose the $\tilde{X}'_1$ coordinate axis of $\tilde{S}'$ to lie along the direction of the normal to the circle at $P(t)$ with its origin coinciding with the origins of $S$ and $S'$ so that its second axis $\tilde{X}'_2$ lies perpendicular to the first axis as indicated in the diagram. $\tilde{S}'$ is an inertial frame moving in a fixed direction with a constant velocity relative to the inertial frame $S$. Thus at time $t$, it simulates the velocity of the rotating disc both in magnitude and direction relative to the frame $S$. It can be used to give a local time variable $t'(t)$ to the disc at the point $P(t)$ at time $t$. The inertial frame $\tilde{S}'$ can acquire it own time keeping by being paired with another inertial frame $\tilde{S}$ rigidly attached to $S$ with its $\tilde{X}_1$ axis parallel with that of $\tilde{X}'_1$ and separated from it by the distance $P(t)L(0) = v(t)t = r(t)\omega t$, the distance that the frame $\tilde{S}'$ has separated from the frame $\tilde{S}$ in the time interval $0 \rightarrow t$, assuming that the angle of frame rotation separation over the same time is given by $\phi (t) = \omega t$. This same distance is represented on the diagram by the arc $P(t)D(0)$. So far the construction has installed a local time keeping from the inertial frame $S$ to the rotating frame $S'$ at the simulation point $P(t)$ which is in effect any point along the path $P(t)$. This is the basis of the formulae (\ref{eq:1.6}) and (\ref{eq:2.8}). The lorentz transformation connecting the two inertial frames of reference frames $\tilde{S}$ and $\tilde{S}'$ has the form 
\begin{eqnarray} \tilde{x}'_1 (t) &=& \tilde{x}_1 (t) \label{eq:0.1} \\
\tilde{x}'_2 (t) &=& \gamma (v(t))( \tilde{x}_2(t)- v(t)t)\label{eq:0.2} \\
t'(t) &=& \gamma (v(t))(t - v(t) \tilde{x}_2(t))/c^2 ) \label{eq:0.3} \\
\gamma (v) &=& (1 -  (v/c)^2)^{-1/2} \label{eq:0.4} \\
v(t) &=& r(t) \omega\label{eq:0.5}
\end{eqnarray}
because the relative motion of these two frames is along the $\tilde{X}'_2$ direction with magnitude $v(t) = r(t)\omega$. As the frame $\tilde{S}'$  was chosen so that $P(t)$  always lies on the $\tilde{X}'_1$ axis (\ref{eq:0.3}) reduces to (\ref{eq:1.6}). Equation (\ref{eq:0.1}) is the hidden assumption in special relativity that the space contraction factor  $\gamma (v)$ applies in full and only to the dimension of the direction of the velocity vector which here is the $\tilde{x}'_2 (t)$ direction. I shall drop this assumption in the second possible rotation transformation (\ref{eq:2.1} $\rightarrow$ \ref{eq:2.9}).

Material related to this general area of ideas can be found in references (\cite{Wol:1958}, \cite{Abr:1990 }, \cite{Dov:1980}, \cite{Cav:1987}, \cite{Gro:1975}, \cite{Ber:1942})

\newpage
\onecolumn
\section{Relativistic Rotation Simulation}
\pagestyle{headings}

\centerline{ \xy (45,-50)*+{\ellipse<195pt>{.}};
\POS(90,-100)*+{0}
\ar @{->} (90,-20)*+{X_2}
\ar @{->} (90,-174)*+{-\bf{e}_2}
\ar @{->} (117,-168)*+{-\bf{e}'_2}
\ar @{->} (21,-128)*+{-\bf{e}'_1}
\ar @{->} (60,-22)*+{X'_2}
\ar @{->>} (130,-40)*+{{\bf P}(t)}
\ar @{.} (147,-62)*+{*}
\ar @{->} (171,-67)*+{X'_1}
\ar @{->} (170,-100)*+{X_1}
\ar @{->} (15,-100)*+{-\bf{e}_1}
\ar @{.} (115,-115)*+{}
\ar @{.>} (74,-90)*+{\tilde X_2}
\POS(113,-115)*+{}
\ar @{.>} (165,-38)*+{\tilde X_1}
\POS(152,-59)*+{}
\ar @{.>} (99,-24)*+{{\bf v}(t)}
\POS(152,-59)*+{}
\ar @{.} (169,-71)*+{L_1}
\POS(155,-59)*+{*\ L(0)}
\ar @{} (134,-20)*+{R_3\ *}
\ar @{} (95,-78)*+{R_4\ *}
\POS(140,-62)*+{D(0)}
\POS(115,-116)*+{\tilde 0}
\POS(132,-37)*+{}
\ar @{.>} (142,-22)*+{\tilde X'_1}
\POS(128,-43)*+{}
\ar @{-} (143,-81)*+{N'_1}
\ar @{-} (72,-63)*+{N'_2}
\ar @{>} (100,-53)*+{}
\ar @{-} (128,-102)*+{N_1}
\ar @{-} (87,-43)*+{N_2}
\POS(127,-40)*+{}
\ar @{->} (120,-24)*+{v(t)cos(\chi(t))}
\POS(114,-48)*+{v(t)sin(\chi(t))}
\POS(51,-61)*+{\omega}
\POS(53,-55)*+{};\POS(45,-65)*+{}
**\crv{(46,-60)} ?>*\dir{>}
\POS(101,-98)*+{\phi(t)}
\POS(105,-101)*+{};\POS(103,-93)*+{}
**\crv{(107,-97)} ?>*\dir{>}
\POS(112,-89)*+{\theta(t)}
\POS(118,-101)*+{};\POS(106,-73)*+{}
**\crv{(118,-85)} ?>*\dir{>}
\POS(114,-76)*+{\chi(t)}
\POS(124,-87)*+{};\POS(110,-67)*+{}
**\crv{(121,-75)} ?>*\dir{>}
\POS(136,-53)*+{\chi(t)}
\POS(133,-57)*+{};\POS(142,-51)*+{}
**\crv{(141,-58)} ?>*\dir{>}
\POS(120,-70)*+{\phi(t)}
\POS(127,-76)*+{};\POS(115,-61)*+{}
**\crv{(124,-65)} ?>*\dir{>}
\POS(110,-64)*+{r(t)}
\POS(90,0)*+{\  Relativistic\ Rotating\ Frame\ Diagram}
\endxy }
\vskip 0.3 cm
The transformation (\ref{eq:0.1} $\rightarrow$ \ref{eq:0.5}) can be used to complete the explanation of the simulation construction as follows. The point $L_1$ has coordinates $( r(t), -(r(t)\tan(\chi (t)) - v(t) t )$ in $\tilde{S}$ and in $\tilde{S}'$ it has coordinates $(r'(t), - r'(t)\tan(\chi '(t))$, denoting the view of the angle $\chi$ in $S'$ as $\chi '$. If we substitute the second components of these into (\ref{eq:0.2}), we obtain, $- r'(t)\tan(\chi '(t))=\gamma (v(t))( -(r(t)(\tan(\chi (t)) $. However from equation (\ref{eq:0.1}) and the diagram $r'(t)=\tilde{x}'_1 (t) = \tilde{x}_1 (t) = r(t)$. It follows that $\tan(\chi '(t))= \gamma (v(t))\tan(\chi (t)$
which re-expressed is equation (\ref{eq:1.4}) of the rotation equation set.
We now only need note from the diagram that $ x'_1 (t) = r (t)\cos (\chi ' (t)) $ and $ x'_2 (t) = r (t)\sin (\chi ' (t)) $.    Thus all the parts of the first transformation have been derived from the rotation {\it linear\/} simulation idea.

\newpage
\subsection{Relativistic Rotation}
\pagestyle{headings}
\centerline{ \xy
\POS(112,-100)*+{L(0)}
\ar @{->} (138,-53.5)*+{R_3}
\ar @{->} (38,-53.5)*+{R_4}
\ar @{->} (112,-43)*+{{\bf v}(t)}
\ar @{->} (38,-100)*+{R_1}
\ar @{->} (138,-100)*+{R_2}
\ar @{->} (138,-68)*+{}
\ar @{->} (38,-68)*+{}
\POS(30,-76)*+{l'_1}
\POS(146,-86)*+{l_1}
\POS(88,-108)*+{l_2}
\POS(75,-50)*+{l_4}
\POS(75,-74)*+{d'_4}
\POS(75,-87)*+{d_4}
\POS(125,-50)*+{l_3}
\POS(125,-72)*+{d'_3}
\POS(125,-87)*+{d_3}
\POS(112,-68)*+{R'_5}
\ar @{->} (138,-68)*+{}
\ar @{->} (38,-68)*+{}
\POS(112,-53.5)*+{R_5}
\ar @{->} (138,-53.5)*+{}
\ar @{->} (38,-53.5)*+{}
\POS(138,-50)*+{ v(t)cos(\chi '(t))}
\POS(138,-100)*+{}
\ar @{.>} (138,-66)*+{\quad\quad R'_3}
\POS(138,-68)*+{*}
\ar @{.>} (138,-54)*+{}
\POS(38,-100)*+{}
\ar @{.>} (38,-66)*+{R'_4 \quad\quad}
\POS(38,-68)*+{*}
\ar @{.>} (38,-54)*+{}
\POS(38,-50)*+{v(t)sin(\chi '(t))}
\POS(118,-76)*+{\chi'(t)}
\POS(111,-71)*+{}; \POS(126,-76)*+{}
**\crv{(120,-69)} ?>*\dir{>}
\POS(120,-65)*+{\chi(t)}
\POS(111,-61)*+{}; \POS(136,-73)*+{}
**\crv{(123,-59)} ?>*\dir{>}
\POS(105,-90)*+{\pi /2}
\POS(95,-91)*+{}; \POS(117,-92)*+{}
**\crv{(105,-83)} ?>*\dir{>}
\POS(90,0)*+{\  Contraction\ Projections\ Diagram}
\endxy }
\vskip 0.3 cm

The second rotation equation is obtained with the same simulation process so that the transformed time $t'(t)$ comes out the same as with the first equation but the dilatation factor, $\gamma(v(t))$, is distributed over the two directions differently. This scheme can be described using the second diagram {\it Contraction Projections Diagram\/}. This scheme is not derived directly from the usual Lorentz transformation but using the first diagram which is non-specific it can be assumed that the transformed geometry can be expected to involve changes of the radial distance $r'(t)$ so that the equality $(r'(t) = r(t))$ does not hold. Such a situation could reasonable hold in the case of the rotating elastic sphere for example. The second formula can be found by using the second diagram which is an enlarged  representation of  a rectangular region of the rotating plane. To see this in place, the diagram should be thought of as reduced in size and then the common points $R_4$, $R_3$ and $L(0)$ of the two diagrams should be brought into coincidence. Thus we have a rectangular region of $S'$ instantaneously in motion in the direction given by the velocity vector ${\bf v}(t)$ so that the Lorentz contraction of the width rest length $l'_1$ is given by equation (\ref{eq:0.8}) and, as can be read off the diagram, the length contraction in the two perpendicular directions of the rectangular coordinates in the frame $S'$ are given by equations (\ref{eq:0.9}) and (\ref{eq:0.10}) below, Thus if we distribute the contractions over the dimensions according to this prescription, we obtain the second set of rotation transformation equations (\ref{eq:2.1} $\rightarrow$ \ref{eq:2.9}).  
\begin{eqnarray}l_1  &=& l'_1(1-(v/c)^2)^{1/2} \label{eq:0.8} \\
d_4  &=& d'_4 (1 - (v\sin (\chi ')/c)^2)^{1/2} = d'_4 \gamma(v_s)^{-1}  \label{eq:0.9} \\
d_3  &=& d'_3 (1 - (v\cos (\chi ')/c)^2)^{1/2} = d'_3 \gamma(v_c)^{-1}. \label{eq:0.10} 
\end{eqnarray}

\twocolumn
The relativistic transformation I shall use in analysing the centrifugal force problem has the form,
\begin{eqnarray}x'_1 (t) &=& r (t)\cos (\chi ' (t)) \label{eq:1.1} \\
x'_2 (t) &=& r (t)\sin (\chi ' (t)) \label{eq:1.2} \\
\gamma (v) &=& (1 -  (v/c)^2)^{-1/2} \label{eq:1.3} \\
\chi '(t) &=& \tan ^{-1} (\tan (\chi (t))\gamma (r(t)\omega)) \label{eq:1.4}\\
\chi (t) &=&\theta (t) - \omega t\label{eq:1.5}\\
t'(t) &=& \gamma ^{-1} (r(t) \omega )t\label{eq:1.6}\\
r'(t) &=& (x'_1(t)^2 + x'_2(t)^2)^{1/2} = r(t).\label{eq:1.7}
\end{eqnarray}

The second relativistic transformation mentioned above but not to be used in this article has the form,
\begin{eqnarray}x'_1 (t) &=& r (t)\cos (\chi(t)) \gamma(v_s(t)) \label{eq:2.1}\\
x'_2 (t) &=& r (t)\sin (\chi (t)) \gamma(v_c(t)) \label{eq:2.2}\\
\gamma (v) &=& (1 -  (v/c)^2)^{-1/2}\label{eq:2.3}\\
v_s(t) &=& r(t) \omega \sin (\chi (t)) \label{eq:2.4} \\
v_c(t) &=& r(t) \omega \cos (\chi (t)) \label{eq:2.5} \\
\chi '(t) &=& \tan ^{-1} (\tan (\chi (t))\frac{\gamma (v_c(t))}{\gamma (v_s(t))}) \label{eq:2.6}\\
\chi (t) &=&\theta (t) - \omega t\label{eq:2.7}\\
t'(t) &=& \gamma ^{-1} (r(t) \omega )t\label{eq:2.8}\\
r'(t) &=& (x'_1(t)^2 + x'_2(t)^2)^{1/2}.\label{eq:2.9}
\end{eqnarray}

The two transformations above (\ref{eq:1.1} $\rightarrow$ \ref{eq:1.7}) and (\ref{eq:2.1} $\rightarrow$ \ref{eq:2.9}) are constructed so as to be from an inertial frame $S$ to a frame $S'$ rotating with an angular velocity $\omega$ relative to the inertial frame $S$. The transformations are not just expressed in terms of the coordinate change from one frame of reference to another because their expression requires the input of some definite path $(r(t),\theta (t))$ which is {\it usually\/} to be taken to be a straight line in space traced out at a constant speed against time. In other words, a path that any free particle should follow according to Newton's first law in the inertial frame $S$. If any other path {\it not\/} conforming to this condition in $S$ is encountered it would imply that the particle is either not free or the inertial character of the frame $S$ is somehow compromised by being in a state of rotation itself, for example. The representation of the two transformations may strike the reader as {\it number\/} redundant but there are so many equations because both collections of formulae include the transformed version in rectangular Cartesian and polar forms and, of course, the extra time variable $t'$ relation to $t$ for the transformed system. 

I shall make use of a specific {\it straight space-line\/} test path in $S$ defined by,

\begin{eqnarray}r(t) &=& (b^2 + (v_b t)^2)^{1/2}\label{eq:3.1}\\
\theta (t) &=& \tan ^{-1}(v_b t/b).\label{eq:3.2}
\end{eqnarray}
The formula for centrifugal acceleration can be derived from general relativity in a number of different ways, see for example Rindler's book \cite{Rin:2001}, and  has the form,
\begin{eqnarray} d^2 r'(t')/dt^{\prime 2} &=& v_{tr}^2 \gamma^2 (v_{tr})/r'(t'). \label{eq:4.1}\end{eqnarray}
I shall take it that this formula is a very reliable result from general relativity theory and use it as an interpretational guide.
The centrifugal acceleration given by equation (\ref{eq:4.1}) only differs from the classical expression by the relativity factor $\gamma^2 (v_{tr})$ where $v_{tr}$ is the polar transverse component of the vector velocity $d{\bf r}'(t')/dt'$. The primes on the various quantities indicate that they are measured by an observer fixed on the rotating frame $S'$ as also is the un-primed $v_{tr}$.
We can make a very significant point about centrifugal force by using the linear test path (\ref{eq:3.1}), (\ref{eq:3.2}) alone and without any reference to relativity as follows. To that end we recall that the polar radial and transverse components, $(a_{ra},a_{tr})$,  of acceleration $a(t)$ in two dimensional motion are given by
\begin{eqnarray} (\ddot{r} -r \dot{\theta}^2, r \ddot{\theta} + 2 \dot{r}\dot{\theta}). \label{eq:5.1} \end{eqnarray}
It follows from the form (\ref{eq:5.1}) that, if a particle is in motion at a constant speed and in a straight line  perpendicular to the radius vector at any instant so that its radial acceleration, $a_{ra}(t)$, is zero, the quantity $\ddot{r}(t)$ takes the place of radial acceleration in the rotating frame and, further, it is equal to $r(t) \dot{\theta} (t)^2$.
Thus forgetting relativity for the moment, we can calculate the value of $\ddot{r}(t)$ directly using the straight line test path (\ref{eq:3.1}), (\ref{eq:3.2})  and evaluate its value at time $t=0$. We find that by giving the velocity parameter $v_p$ the value $\omega b$ and direct calculation that    
\begin{eqnarray} (\ddot{r}(0) = r(0) \dot{\theta}(0)^2 = b \omega^2. \label{eq:5.2} \end{eqnarray}
This is the usual formula for the centrifugal acceleration experience by a particle fixed to a disc at a distance $b$ from its centre and which is 
rotating at angular velocity $\omega$. However, the particle on this path is moving in a straight line with no connection to any rotating reference frame. All that has been done here is to give the constant velocity test path a distance b from the origin at time $t=0$ the same value that the rotating disc has at the radial distance b from its centre of rotation. Thus a straight line non-accelerating path passing perpendicularly to a radius of the disc and instantaneously having the velocity of the disc at the point of least distance from the origin appears from the disc view to have a centrifugal acceleration of the exactly correct classical form.
This certainly means that the form of the centrifugal acceleration formula is determined only by {\it kinematics\/}, there is no dynamics involved and so classically we must conclude that there is no obvious link between centrifugal acceleration and inertia. This reinforces the generally held idea that centrifugal force is a {\it fictitious\/} force. It seems to me that force only comes into this story when a definite valued mass m is to be held attached at a fixed radial distance from the origin of a rotating object. The {\it true force\/} then needed arises at least classically as the {\it centripetal\/} force $f_{cp} =- m a_{ra}(t)= -f_{cf}$, the negative of the centrifugal force discussed above. Clearly, for a particle to be held in a circular orbit, the free particle tendency to travel in a straight line at constant speed must be overcome. Of course this is just the orthodoxy and it appears from the point of view taken here to be correct.
$ f_{cp}=-a_{ra}(t)$ is measured in the rotating frame but here the force is determined by the mass rather than the inertial mass being determined by some mysterious influence of other masses in the vicinity or far away. We can use the new transformation ((\ref{eq:1.1} $\rightarrow$ \ref{eq:1.7})) to check whether relativity gives the same conclusion. However, the last few sentences above do seem to leave some small uncertainty about the relation between distant and local mass values.  Perhaps someone can produce a more definitive case one way or the other.   

\section{Relativistic Centrifugal\newline Acceleration}
\setcounter{equation}{0}
\label{sec-reLcen}

It is thus imperative that we should check out whether or not the same conclusion follows from a relativistic version of the above argument. In order to carry forward this analysis we need to look at what happens if the base frame $S$, originally assumed to be inertial, is some way compromised and actually contains particle tracks that would indicate that it is actually rotating. We can achieve this by reading from the transformation how the rotating frame $S'$ assumed rotation at some angular velocity value $\omega_0$, say, would see our linear test path. This is achieved by substituting the original test path (\ref{eq:3.1}), (\ref{eq:3.2})  into the transformation and obtaining its image as seen in $S'$. This gives us a new test path called $P' \equiv (r'(t),\chi '(t),t'(t))$, say,  which can now be substituted back as an input into the transformation.  Then the  centrifugal acceleration can be calculated as before but using the non-linear questionable path planted on the assumed inertial frame $S$. This calculated centrifugal acceleration can then be used to see how it has to be changed so as to be consistent with or exactly equal to the known correct formula for centrifugal acceleration from general relativity. 

The question that arises is how dependent is the form of the relativistic centrifugal acceleration on the linearity of our test path formula $P'$.
On carrying out the calculation which involves a full use of the relativistic rotation transformation and consequent re-expressing the transformed variables into their form of dependence on the $t'$ variable we need the results from equations (\ref{eq:6.1}), (\ref{eq:6.2}) and  (\ref{eq:6.3}). The calculation is much simplified by evaluating all the quantities at time $t=0$ which is implied by using the $|_0$ symbol.
We find that the formula involves the $\omega_0$ associated with the questionable path and is the last of,

\begin{eqnarray} d^2 r'(t)/dt^{\prime 2}|_0 &=& r(0) (d\theta '(t)/dt'|_0)^2|\label{eq:6.1}\\
d\theta '(t)/dt'|_0 &=& (d\theta ' (t) /dt|_0) /(dt'(t)/dt|_0)\label{eq:6.2}\\
dt'(t)/dt|_0 &=& (1 - (b \omega /c)^2)^{1/2} \label{eq:6.3}\\
d^2 r'(t)/dt^{\prime 2}|_0 &=& \frac{b(\omega_0 - v_b/b)^2 }{(1 - (b \omega/c)^2)(1 - (b\omega _0/c)^2)}\nonumber\\
\label{eq:6.4}
\end{eqnarray}
and if this is to be consistent with the form that comes from general relativity considerations then taking the moving particle as fixed instantaneously to the $x'_1$ at time $t=0$ at a distance b from the origin of rotation so that $v_b = b\omega$ and also denoting $v_{tr} = b \omega ' $ it is necessary that 

\begin{eqnarray}\frac{{\omega '}^2}{1 - (b\omega '/c)^2} &=& \frac{(\omega_0 - \omega)^2 }{(1 - (b\omega/c)^2)(1 - (b\omega _0/c)^2)}.\nonumber\\
\label{eq:7.1} \end{eqnarray}

Suppose now that the angular velocity $\omega '$ is the {\it apparent\/} angular velocity of of the celestial sphere as observed from a flat disk with its centre rigidly attached to the earth's surface in a tangent plane configuration centred on the earth's axis of rotation. Earth here, can be replaced by any solid rotating platform anywhere.
There are two identifications for $\omega '$ that preserves the simple factor form on both sides of equation (\ref{eq:7.1}) and simultaneously identify a {\it local\/} frame of reference that is in a state of absolute zero rotation itself. These two local frames are the $S$ and $S'$ of the transformation formulae (\ref{eq:1.1}) $\rightarrow$ (\ref{eq:1.7}).  The identifications are $ \omega ' = \omega \Rightarrow \omega_0 = 0$ or $\omega ' = \omega_0 \Rightarrow \omega = 0$. It is easy to check that any other identifications for $\omega '$ destroy this simple factor structure and further are complicated and aesthetically unattractive. Thus can we can make the following interpretation of this result. If the general relativity formula for centrifugal acceleration is accepted as correct throughout the universe then it must be used against and relative to a local background reference frame that is in a local definite absolute state of non-rotation coincident with the state of non-rotation that can be determined by observation of the celestial sphere. Expressed otherwise, there is in general relativity everywhere to be found a local reference frame in a state with absolutely no rotation relative to which any local rotating state will have a definite {\it absolute\/} value of angular velocity.  However, this seems to have nothing very directly to do with inertia. The further very important consequence follows from this section.  If a rigid physical object with a directional arrow marked on it invariably placed with respect to its geometrical form moves through space time either by propulsion or freefall its directional arrow can be kept un-rotating with respect to a local absolute non-rotating local reference frame which is the same thing as keeping it pointing to a fixed identified object on the celestial sphere. Of course this navigation process can be helped by having an active gyroscope with rotation axis coincident with the telescope so as to hold the telescope on its object and here again the gyroscope keeps its constant orientation by reference to the local non-rotating frame.  In fact, this is just stating the physical basis for space navigation in practice. We note also in this context that keeping a distant object in the line of sight of such a marked arrow involves locking a {\it local\/} telescope in line with the {\it local\/} space part of the null geodesic which is the channel along which the information from the distant object is coming. In general relativity, the passage through space-time of a {\it free\/} photon is considered to take place along a null geodesic and in quantum theory such a photon is also considered to carry a unique quantum of {\it vector\/} angular momentum of magnitude $\hbar$ lying exactly in its direction of propagation. Thus the photon in its journey through space-time will always be locked onto the {\it local\/} state of absolute non-rotation. Alternatively expressed, this can be regarded as a process in which the free photon field connects all the local non-rotation states throughout the whole electromagnetic connected universe. A photon {\it exists\/} on the null geodesic between its creation and destruction events because there is zero {\it proper\/} time between these two events.

\section{Conclusions} It seems that Mach's principle is true in the limited sense of the existence of a universal standard of absolute local rotation relating the celestial sphere to local conditions. This does not, of course, preclude the existence of relatively rotating systems with a relative angular velocity. Thus this limited Mach's Principle partially restores Newton's absolute space background in terms of absolute {\it angular\/} velocity while leaving the {\it relative\/} relativistic status of {\it linear\/} velocity unchanged. Thus it is the universal network of actual or potential null geodesics that unite all the local states of absolute rotation. When we observe the apparent rotation of the celestial sphere, we are actually looking at some local mapping of the celestial sphere on some local present time two dimensional representation of the terminal points of null geodesics that originated in the greatly distant past from greatly distant objects. The two dimensional representation at one level could be a telescopic viewing screen  at another level, the retina of the eye or at a deeper level it is an image in the human brain. Always though, we are actually seeing local movement against the local standard of no rotation, actually just the earth's rotation against this local standard.

Finally, I would like to mention my motivation for this study which was work(\cite{Gil:1996}, \cite{Gil:1999}) in quantum mechanics on the {\it fine structure constant\/} problem. I had long been puzzled by the fact that calculations on hydrogen like atom electronic orbits seemed to assume that the nucleus or a centre of gravity was fixed on a true inertial frame and one could get from theory the values of two perfectly definite {\it quantized\/} quantities, the Bohr radii, $\lambdabar _C/(Z\alpha)$, and the velocities, $Z\alpha c$, of the electrons on such orbits. This implies that in such orbits the electron moves with a definite angular velocity, $ m_e(Z\alpha c)^2/\hbar$. It seemed to me that this only made sense if these angular velocities were  {\it absolute\/} as we have been discussing above. Further more, I had managed to derive a very accurate value for the fine structure constant using the equation of centrifugal force. This implies that centrifugal force is true even under condition when the coupling is large, $137\alpha$ approximately unity, as in the case of hydrogen-137, way outside the quantum electrodynamics realm of the {\it small\/} expansion parameter, $\alpha \approx 1/137$. The work described in this article seems to reinforce my view that there is a local standard of zero rotation everywhere.

\section{Acknowledgements}

I am greatly indebted to Professors Clive Kilmister and Wolfgang Rindler for help, encouragement and inspiration over many years.


\begin{thebibliography}{99}
\bibitem{Sos:1980} Soshichi Uchii {at http://www.bun.kyoto-u.ac.jp/\%7Esuchii/mach.pr.html}
\bibitem{Gil:1996}  Gilson, J.G. {1996, Calculating the fine structure constant
, Physics Essays, {\bf  9} , 2 June, 342-353}
\bibitem{Gil:1999}  Gilson, J.G. {1999, The fine structure constant, http://www.fine-structure-constant.org/}   
\bibitem{Rin:2001}  Rindler, W. {2001, Relativity: Special, General and Cosmological, Oxford University Press}
\bibitem{Mac:1893}  Mach, H. {1893, The Science of Mechanics, Chicago, IL: Open Court}
\bibitem{Bar:1995}  Barbour, J. and Pfister, H. {1995, Mach's Principle: From Newton's Bucket to Quantum Gravity. Boston, MA: Birkhäuser}
\bibitem{Mis:1973}  Misner, C. W.; Thorne, K. S.; and Wheeler, J. A. {1973, Gravitation. Boston, San Francisco, CA: W. H. Freeman}
\bibitem{Wol:1958}  Wolfgang Pauli {1958, Theory of Relativity, 130--134, Pergamon Press }
\bibitem{Abr:1990 } Abraham Pais, {1990, Subtle is the Lord: the Science and Life of Albert Einstein, p 214}
\bibitem{Bel:1987} J. S. Bell {1987, Speakable and Unspeakable in Quantum Mechanics, p 67, Cambridge University Press}
\bibitem{Dov:1980} Dover {1980, The Principle of Relativity, Dover}
\bibitem{Cav:1987} G. Cavalleri {1987, Nuovo Cimento 53B, p 415}
\bibitem{Gro:1975} O. Gron {1975,  AJP, Vol. 43,  No. 10, p 869 }
\bibitem{Ber:1942} C. Berenda {1942, Phys. Rev. 62, p 280 }
\bibitem{Wei:1988} M. Weiss {1988,\\ http://www.physics.adelaide.edu.au/$\sim$ dkoks/Faq/Relativity/SR/rigi$d\_$disk.html}

\end{thebibliography}
\enddocument